\documentclass[floats,floatfix,showpacs,amssymb,prd,twocolumn,superscriptaddress,nofootinbib,aps]{revtex4-1}

\usepackage{graphicx, epsf, epsfig, amssymb}% Include figure files
\usepackage{bm}
\usepackage{longtable}
\usepackage[usenames,dvipsnames]{xcolor}
\usepackage{color}
\usepackage[breaklinks]{hyperref}
\usepackage{amsfonts,amsmath,amssymb,mathrsfs}
\usepackage{soul}

\def\be{\begin{equation}}
\def\ee{\end{equation}}
\def\beq{\begin{eqnarray}}
\def\eeq{\end{eqnarray}}
\def\nn{\nonumber}

\begin{document}

\title{Superradiant instability of black holes immersed in a magnetic field}

\author{Richard Brito}
\email{richard.brito@tecnico.ulisboa.pt}
\affiliation{CENTRA, Departamento de F\'{\i}sica, Instituto Superior
  T\'ecnico, Universidade de Lisboa, Avenida~Rovisco Pais
  1, 1049 Lisboa, Portugal}
\affiliation{Perimeter Institute for Theoretical Physics
Waterloo, Ontario N2J 2W9, Canada
}

\author{Vitor Cardoso}
%\email{vitor.cardoso@tecnico.ulisboa.pt}
\affiliation{CENTRA, Departamento de F\'{\i}sica, Instituto Superior
  T\'ecnico, Universidade de Lisboa, Avenida~Rovisco Pais
  1, 1049 Lisboa, Portugal}
\affiliation{Perimeter Institute for Theoretical Physics
Waterloo, Ontario N2J 2W9, Canada
}
\affiliation{Department of Physics and Astronomy, The University of Mississippi, University, MS 38677, USA.
}

\author{Paolo Pani}
%\email{paolo.pani@tecnico.ulisboa.pt}
\affiliation{CENTRA, Departamento de F\'{\i}sica, Instituto Superior
  T\'ecnico, Universidade de Lisboa, Avenida~Rovisco Pais 1, 1049 Lisboa, Portugal}

\begin{abstract}
Magnetic fields surrounding spinning black holes can confine radiation and trigger superradiant instabilities.
To investigate this effect, we perform the first fully-consistent linear analysis of the Ernst spacetime, an exact solution of the Einstein--Maxwell equations describing a black hole immersed in a uniform magnetic field $B$.
In the limit in which the black-hole mass vanishes, the background reduces to the marginally stable Melvin spacetime. The presence of an event horizon introduces a small dissipative term, 
resulting in a set of long-lived -- or unstable -- modes. We provide a simple interpretation of the mode spectrum in terms of a small perfect absorber immersed in a confining box of size $\sim1/B$ and show that rotation triggers a superradiant instability. By studying scalar perturbations of a magnetized Kerr--Newman black hole, we are able to confirm and quantify the details of this instability. The instability time scale can be orders of magnitude shorter than that associated to massive bosonic fields. The instability extracts angular momentum from the event horizon, competing against accretion. This implies that strong magnetic fields set an upper bound on the black-hole spin. Conversely, observations of highly-spinning massive black holes impose an intrinsic limit to the strength of the surrounding magnetic field. We discuss the astrophysical implications of our results and the limitations of the Ernst spacetime to describe realistic astrophysical configurations.
\end{abstract}

\pacs{04.70.-s, 04.25.Nx}
\maketitle

%%%%%%%%%%%%%%%%%%%%%%%%%%%%%%%%%%%%%%%%%%%%%%%%%%%%%%%%%%%%%%%
\section{Introduction}
%%%%%%%%%%%%%%%%%%%%%%%%%%%%%%%%%%%%%%%%%%%%%%%%%%%%%%%%%%%%%%%
The existence of strong magnetic fields around astrophysical black holes (BHs) is believed to be at the origin of some of the most energetic events of our Universe, such as the emission of relativistic jets. The Blandford-Znajek process is widely believed to be one leading mechanism at the origin of these phenomena~\cite{Blandford:1977ds}. This process allows us to extract energy from a spinning BH due to the presence of a magnetic field supported by the material accreted by the BH.

A full comprehension of the interactions between the accretion disk, the surrounding magnetic field and the BH is a complex problem and requires the use of sophisticated general-relativistic magnetohydrodynamic simulations (cf. e.g. Refs.~\cite{2012MNRAS.423.3083M,2013MNRAS.436.3741P}), nonetheless a qualitative picture can be drawn by studying stationary magnetized BH solutions in general relativity. For example, the approximate solution found by Wald~\cite{Wald:1974np}, which describes a Kerr BH immersed in a test uniform magnetic field aligned with the BH spin axis, has served as a model to understand the interaction of BHs with magnetic fields. Several remarkable phenomena --~such as charge induction~\cite{Wald:1974np} and a Meissner-like effect~\cite{PhysRevD.12.3037}~-- can be understood by studying this simple solution~\cite{Penna:2014aza}. 

In addition to the perturbative solution found by Wald, a class of \emph{exact} solutions of the Einstein--Maxwell equations, describing BHs immersed in a uniform magnetic field, was discovered by Ernst, who developed a powerful method to construct them starting from vacuum solutions of Einstein's equations~\cite{Ernst:1976BHM}. Although the Ernst spacetimes are not asymptotically flat, they can also be used to model the properties of BHs immersed in strong magnetic fields in a simple way. 

Even though the Wald and Ernst solutions were discovered 40 years ago, the dynamics of linear perturbations in these backgrounds is still largely unexplored.
One of the main motivations to study perturbations of magnetized rotating BHs is the possibility, first proposed by Galt'sov and Petukhov~\cite{Galtsov:1978ag}, that the magnetic field can trigger superradiant instabilities~\cite{Press:1972zz,Cardoso:2013krh}. 
To occur, superradiant instabilities need essentially two ingredients: (i) a monochromatic bosonic wave with low-frequency $\omega$ satisfying the superradiance condition,
\be
\omega<m\Omega_H\,,
\ee
where $\Omega_H$ is the angular velocity of the BH horizon and $m$ is an integer characterizing the azimuthal dependence of the wave; and (ii) a mechanism to trap superradiant modes near the BH. The first condition allows to extract rotational energy from the BH, spinning it down, while the second condition is necessary to ``keep the extraction going,'' thereby triggering the instability.

Several confining mechanisms to trap the modes have been investigated, starting from an artificial mirror around the BH (the so-called ``BH bomb''~\cite{Press:1972zz,Cardoso:2004nk,Cardoso:2013krh}), to more natural ones like massive bosonic fields~\cite{Damour:1976kh,Zouros:1979iw,Detweiler:1980uk,Hod:2012px,Herdeiro:2014goa}, where the mass term plays the role of the mirror, or the asymptotically anti-de Sitter (AdS) spacetime, where the AdS boundary confines the perturbations inside the bulk~\cite{Cardoso:2004hs,Cardoso:2006wa,Dias:2013sdc,Cardoso:2013pza}.
For the case of massive bosonic fields the instability has been studied extensively for scalar~\cite{Damour:1976kh,Zouros:1979iw,Detweiler:1980uk,Furuhashi:2004jk,Cardoso:2005vk,Dolan:2007mj,Hod:2012px,Herdeiro:2014goa}, vector~\cite{Rosa:2011my,Pani:2012bp,Pani:2012vp,Witek:2012tr} and tensor~\cite{Brito:2013wya} fields. 

Magnetic fields can confine the radiation in a similar way. Working in a $Br\ll 1$ expansion (with $B$ being the magnetic field strength and $r$ the radial coordinate, both in geometric units), Refs.~\cite{Galtsov:1978ag,Konoplya:2007yy,Konoplya:2008hj} showed that a scalar field propagating on the Ernst background is equivalent to a massive scalar perturbation propagating on a Schwarzschild or Kerr metric with an effective mass $\mu_{\rm eff}=Bm$. As such, the magnetic field triggers the same superradiant instability associated to massive fields. However, such approximation becomes inaccurate at distances comparable to or larger than $\sim 1/B$. As we show, this profoundly affects the dynamics of the perturbations, because the spectrum is defined by physically-motivated boundary conditions imposed at large distances $r\gg 1/B$.

In this work we take a step further to understand how strong magnetic fields affect BH spacetimes. We study scalar perturbations of the Ernst solutions with no approximation for the first time. We show that magnetized BHs can indeed support superradiant unstable modes and that this instability can be orders of magnitude stronger than the one estimated using the approximation of Refs.~\cite{Galtsov:1978ag,Konoplya:2007yy,Konoplya:2008hj} in terms of an effective mass $\mu_{\rm eff}=Bm$. In the exact case the perturbation equations do not seem to be separable and this prevents the use of most methods to compute quasinormal modes (QNMs) (see~\cite{Kokkotas:1999bd,Nollert:1999ji,Berti:2009kk} for reviews). In this work we circumvent this problem using powerful techniques developed in the past few years (see e.g.~\cite{Pani:2013pma}), which allow us to solve the full linearized dynamics for any value of $B$.

%%%%%%%%%%%%%%%%%%%%%%%%%%%%%%%%%%%%%%%%%%%%%%%%%%%%%%%%%%%%%%%
\subsection{Executive summary and plan}
%%%%%%%%%%%%%%%%%%%%%%%%%%%%%%%%%%%%%%%%%%%%%%%%%%%%%%%%%%%%%%%

We find that the spectrum of unstable modes of spinning Kerr--Newman magnetized BHs~\cite{Aliev:1988wy,Gibbons:2013yq} is different from that of massive scalar fields on a vacuum Kerr spacetime, and it is instead analogous to the BH bomb case~\cite{Press:1972zz,Cardoso:2004nk,Cardoso:2013krh}.
In fact, since the solutions found by Ernst are not asymptotically flat, there is no reason to believe that superradiant instabilities in this spacetime resemble the ones triggered by massive fields, even for arbitrarily small magnetic field.

At infinity the background resembles a solution of the Einstein--Maxwell describing a uniform magnetic field held together by its own gravitational pull. This solution, which was found by Melvin~\cite{Melvin:1963qx,Melvin:1965zza} and further studied by Thorne~\cite{PhysRev.139.B244}, is known in the literature as the Melvin spacetime. Like the AdS spacetime, the Melvin solution admits normal modes (computed in Sec.~\ref{sec:Melvin}), because the asymptotic boundary of the Melvin solution acts as a confining box for perturbations. Once a BH is added to the spacetime, absorption or amplification at the horizon is possible, as happens for a small BH immersed in AdS spacetime~\cite{Horowitz:1999jd,Cardoso:2001bb}. We verify this argument by computing the modes explicitly for
the Schwarzschild--Ernst solution in Sec.~\ref{sec:Ernst}.

As we argue in Sec.~\ref{sec:model}, it is possible to describe
{\it any} confined BH geometry using just the low-frequency absorption cross section of the (isolated, in flat spacetime) BH. 
We show this reproduces known results in the literature.
This simple model predicts that superradiant instabilities arise when the BH rotates, a prediction that we verify also explicitly in Sec.~\ref{sec:Ernst_rot}, where we study rotating magnetized BHs.
Finally, we discuss the astrophysical implications of our results and close with some concluding remarks.

We use $G=c=4\pi\epsilon_0=1$ units, where $G$, $c$ and $\epsilon_0$ are the Newton constant, the speed of light and the  vacuum permittivity, respectively.

%%%%%%%%%%%%%%%%%%%%%%%%%%%%%%%%%%%%%%%%%%%%%%%%%%%%%%%%%%%%%%%%%%%%%%%%%%%%%%%%%
\section{The Melvin spacetime and its normal oscillation modes}\label{sec:Melvin}
%%%%%%%%%%%%%%%%%%%%%%%%%%%%%%%%%%%%%%%%%%%%%%%%%%%%%%%%%%%%%%%%%%%%%%%%%%%%%%%%%
We start by studying the geodesic motion and the normal modes of the Melvin spacetime, which are propaedeutic to understand the QNMs of a BH immersed in a magnetic field. In cylindrical coordinates the Melvin metric is given by~\cite{Melvin:1963qx}
\be
\label{Melvin}
ds^2=\Lambda^2\left(-dt^2+d\rho^2+dz^2\right)+\frac{\rho^2}{\Lambda^2}d\phi^2\,,
\ee
where $\Lambda=1+B^2\rho^2/4$. This solution describes a uniform magnetic field aligned along the $z$-axis. 

Let us start with a brief geodesic analysis of the metric~\eqref{Melvin}. Staticity and axial symmetry of the metric imply the existence of a conserved energy $E$ and angular momentum parameter $L$, defined as
\begin{equation}
\Lambda^2\dot{t}=E\,, \qquad \frac{\rho^2}{\Lambda^2}\dot{\phi}=L\,,
\end{equation}
where a dot stands for derivative with respect to an affine parameter. Null particles then obey the equation
\be
\dot{\rho}^2=V_\rho\equiv \frac{E^2}{\Lambda^4}-\frac{L^2}{\rho^2}\,,
\ee
or simply
\be
\left(\frac{d\rho}{dt}\right)^2=1-\frac{\Lambda^4\left(L/E\right)^2}{\rho^2}\,.
\ee
Circular ($V_\rho=dV_\rho/d\rho=0$) geodesics for massless particles are only possible for $\rho^2=4/(3B^2)$ and correspond to an angular frequency 
\be
\Omega\equiv \frac{d\phi}{dt}=\frac{16\sqrt{3}\,B}{9\sqrt{4}}\sim 1.5396 B \label{eq:nullgeodesics}\,.
\ee

In the geometric-optics regime, normal modes with $m\gg 1$ in the Melvin spacetime are expected~\cite{Cardoso:2008bp} to reduce to the geodesic result described by Eq.~\eqref{eq:nullgeodesics}, i.e,
\be
\omega_{\rm normal}=m\Omega=\frac{16\sqrt{3}\,mB}{9\sqrt{4}}\sim 1.5396\,mB\,.\label{eq:nullgeodesics2}
\ee

Let us now find the normal modes of a probe scalar field propagating in the Melvin metric~\eqref{Melvin}. The Klein-Gordon equation for a massless field has the form
\be\label{KG}
\Box\Phi\equiv\frac{1}{\sqrt{-g}}\left(g^{\mu\nu}\sqrt{-g}\Phi_{;\mu}\right)_{;\nu}=0\,.
\ee
By making the following ansatz for the scalar field
\be
\Phi(t,\rho,z,\phi)=\frac{Q(\rho)}{\sqrt{\rho}} e^{i k z}e^{i m \phi} e^{-i \omega t}\,,
\ee
the Klein-Gordon equation \eqref{KG} reads
\be\label{melvinKG}
Q''(y)+\left[{\tilde{\omega}}^2-\frac{m^2 \left(y^2+4\right)^4-64}{256 y^2}\right]Q(y)=0\,,
\ee
where ${\tilde{\omega}}^2=(\omega^2-k^2)/B^2$ and $y=B\rho$. Note that with these redefinitions the magnetic field $B$ scales out of the problem. Furthermore, the eigenvalue problem is invariant under $m\to -m$, $\omega\to -\omega$, $k\to -k$ and $Q(y)\to Q^*(y)$ so we consider only modes with $m>0$.

After imposing appropriate boundary conditions, Eq.~\eqref{melvinKG} defines a boundary value problem that admits normal modes. Near the origin the solution behaves as
\begin{equation}\label{or_melvin}
 Q(y)\sim A_1 y^{m+1/2}+A_2 y^{-m+1/2}\,,
\end{equation}
%%%
and regularity at the origin imposes $A_2=0$. The asymptotic behavior at infinity is given by
\begin{equation}
 Q(y)\sim y^{-3/2}\left[C e^{y^4m/64}+ D e^{-y^4m/64}\right]\,, \label{inf_melvin}
\end{equation}
and the only acceptable physical solution corresponds to $C=0$. 

To find the normal frequencies of this spacetime we integrate numerically Eq.~\eqref{melvinKG} starting from the boundary condition~\eqref{or_melvin} and imposing $C=0$ in the asymptotic solution~\eqref{inf_melvin}. This selects a discrete spectrum of frequencies which are summarized in Tables~\ref{tab:spectrum_melvin1} and~\ref{tab:spectrum_melvin2}.
\begin{table}[t]
% \scriptsize
\centering \caption{Scalar normal modes of the Melvin spacetime for $m=1$ and different overtone number $n$.} 
\vskip 12pt
\begin{tabular}{@{}ccccccccccccccccccccc@{}}
\hline \hline
\multicolumn{7}{c}{${\tilde{\omega}}$}\\ 
\hline
$n=0$      & $n=1$   & $n=2$    & $n=3$   &$n=4$    & $n=5$  & $n=6$\\
\hline 
2.04862    & 2.91334 & 3.68457  & 4.39629	&5.06541  &5.70187 & 6.31212\\
\hline \hline
\end{tabular}
\label{tab:spectrum_melvin1}
\end{table}

\begin{table}[t]
% \scriptsize
\centering \caption{Fundamental ($n=0$) scalar normal modes of the Melvin spacetime for different azimuthal number $m$.} 
\vskip 12pt
\begin{tabular}{@{}ccccccccccccccccccccc@{}}
\hline \hline
\multicolumn{6}{c}{${\tilde{\omega}}$}\\ 
\hline
$m=1$      & $m=2$   & $m=3$    & $m=4$   &$m=5$    & $m=6$  \\
\hline 
2.04862    & 3.59874 &5.14195   &	6.68336 &8.22404  &	9.76436\\
\hline \hline
\end{tabular}
\label{tab:spectrum_melvin2}
\end{table}
The most important points to retain from these results are that (i) Melvin spacetimes are (marginally) stable and are described by a set of {\it normal} modes; (ii)
for large $m$ our results are well consistent with the expansion ${\tilde{\omega}}=1.5396m+ 0.5301-0.02113/m$, in excellent agreement with the geodesic analysis in Eq.~\eqref{eq:nullgeodesics2}.

At this point it is important to stress that these modes only exist due to the behavior of the metric at $\rho\to\infty$, which is not asymptotically flat. Indeed, considering $y\equiv B\rho\ll 1$ and neglecting terms at $\mathcal{O}(y^2)$, we find that Eq.~\eqref{melvinKG}  describes a scalar field propagating in Minkowski spacetime with effective mass $\mu_{\rm eff}=mB$. A free massive field in flat space does not form stationary bound states and, therefore, no modes are predicted for the Melvin spacetime within this approximation. These modes solely exist due to the boundary condition~\eqref{inf_melvin} imposed by the magnetic field at large distances, $B\rho\gg1$. The situation is analogous to what happens in AdS spacetime. Normal modes exist in pure AdS space due to the timelike boundary at spatial infinity, which allows null rays to reach the boundary in a finite time and be reflected back. In this case the AdS radius selects the frequencies of these modes. In the same way, in the Melvin spacetime perturbations are confined by the magnetic field which behaves like an infinite ``wall'' at a radius $r_0\sim 1/B$. This allows for the existence of a discrete set of normal modes. As we discuss in the next sections, such modes would be missed by a perturbative analysis similar to what was done in Refs.~\cite{Galtsov:1978ag,Konoplya:2008hj,Konoplya:2007yy}, where QNMs of a BH immersed in the Melvin universe have been computed perturbatively to order $B^2$.

%%%%%%%%%%%%%%%%%%%%%%%%%%%%%%%%%%%%%%%%%%%%%%%%%%%%%%%%%%%%%%%%%%%%%%%%%%
\section{The linear stability of the Ernst spacetime}\label{sec:Ernst}
%%%%%%%%%%%%%%%%%%%%%%%%%%%%%%%%%%%%%%%%%%%%%%%%%%%%%%%%%%%%%%%%%%%%%%%%%%
%%%%%%%%%%%%%%%%%%%%%%%%%%%%%%%%%%%%%%%%%%%%%%%%%%%%%%%%%%%%%%%%%%%
\subsection{The Ernst background spacetime}
%%%%%%%%%%%%%%%%%%%%%%%%%%%%%%%%%%%%%%%%%%%%%%%%%%%%%%%%%%%%%%%%%%%
In 1976 Ernst found a class of exact BH solutions of the Einstein--Maxwell equations immersed in the Melvin spacetime~\cite{Ernst:1976BHM}. The simplest of these solutions corresponds to a magnetized Schwarzschild BH, also known as the Ernst metric, which is given by
\be\label{Ernst}
ds^2=\Lambda^2\left(-f(r)dt^2+\frac{dr^2}{f(r)}+r^2d\theta^2\right)+\frac{r^2\sin\theta^2}{\Lambda^2}d\phi^2\,,
\ee
where $f(r)=1-\frac{2M}{r}$. In the limit $M\to 0$ this metric reduces to the Melvin solution~\eqref{Melvin}, with $\rho=r \sin\theta$ and $z=r \cos\theta$, while in the limit $B\to 0$ it reduces to the standard Schwarzschild solution. Due to the presence of the magnetic field this spacetime is not asymptotically flat, but instead approaches the Melvin metric as $r/M\to\infty$. The vector potential giving rise to the homogeneous magnetic field reads
\be
A_{\mu}dx^{\mu}=-\frac{B r^2 \sin^2\theta}{2\Lambda}d\phi\,.
\ee

The event horizon is located at $r_H=2M$ and its area is given by $A_H=4\pi r_H^2$, as in the Schwarzschild BH, but due to the $\theta$--dependence of the $g_{\phi\phi}$ component, the horizon takes the form of a cigar-shaped object~\cite{Wild:1980zz}. However the magnetic field only starts to distort significantly the spacetime at distances of the order of $B^{-1}$.

%%%%%%%%%%%%%%%%%%%%%%%%%%%%%%%%%%%%%%%%%%%%%%%%%%%%%%%%%%%%%%%%%%%
\subsection{Linearized analysis}
%%%%%%%%%%%%%%%%%%%%%%%%%%%%%%%%%%%%%%%%%%%%%%%%%%%%%%%%%%%%%%%%%%%
In a Melvin background the scalar field equation can be separated using cylindrical coordinates. However, due to the presence of the BH, cylindrical symmetry is lost in the Ernst metric, making the separation of the radial and angular part apparently impossible. Nevertheless we can use the method discussed in~\cite{Dolan:2012yt} to separate the equation at the expense of introducing couplings between different modes (see also Ref.~\cite{Pani:2013pma} for a review).

We begin by splitting the angular and radial dependence of the field as
\be
\Phi(t,r,\theta,\phi)=\sum_{jm}\frac{Q_j(r,t)}{r}Y_{jm}(\theta,\phi)\,,
\ee
where $Y_{jm}(\theta,\phi)$ denotes the usual spherical harmonics. 
Because the background is axisymmetric, the eigenfunctions are degenerate in the azimuthal number $m$.
Inserting the ansatz above in the Klein-Gordon equation~\eqref{KG} and considering the background \eqref{Ernst}, we find
\be\label{KG_ernst_1}
\sum_{jm} Y_{jm}(\theta,\phi)\left[\frac{d^2Q_j}{dr^2_*}-\frac{d^2Q_j}{dt^2}-V_{\rm eff}(r,\theta) Q_j\right]=0\,,
\ee
where $r_*$ is the tortoise coordinate, defined via $dr/dr_*=f$, and
\beq\label{potential}
V_{\rm eff}&&(r,\theta)=f(r)\left\{\frac{j(j+1)}{r^2}+\frac{2M}{r^3}\right.\nn\\
&& +\frac{B^2 m^2}{256}\left[\left(B^2 r^2+8\right)
   \left(B^4 r^4+8 B^2 r^2+32\right) \right.\nn\\
&&\left.- B^2 r^2 \left(3 B^4 r^4+32 B^2 r^2+96\right) \cos ^2{\theta}\right.\nn\\
&&\left. +B^4 r^4 \left(3 B^2 r^2+16\right) \cos ^4\theta\left.  -B^6 r^6 \cos ^6\theta \right]\right\}\,.
\eeq

Since the spacetime is not spherically symmetric, the angular and radial parts of the Klein-Gordon equation cannot be separated using a basis of spherical harmonics. Nonetheless, the problem can be reduced to a $1+1$--problem using the fact that terms with $\cos^n\theta$ lead to couplings between different multipoles~\cite{Dolan:2012yt}. To show this, we first multiply Eq.~\eqref{KG_ernst_1} by $Y_{lm}^*(\theta,\phi)$ and integrate over the sphere. Then, making use of the fact that the Clebsch-Gordan coefficients,
\be
c^{(n)}_{jlm}\equiv\left\langle l m\left|\cos^n\right|j m\right\rangle\,,
\ee
are zero unless $j=l$ or $j=l-n,....,l+n$,
we finally arrive at the following equation
\begin{align} \label{KG_ernst}
&\frac{d^2Q_l(r,t)}{dr^2_*}-\frac{d^2Q_l(r,t)}{dt^2}-\sum_{i=-3}^{3}V_{l+2i} Q_{l+2i}(r,t)=0\,,
\end{align}
where the radial potentials read
%
% \begin{widetext}
%
\beq
V_{l}&&=f\left\{\frac{l(l+1)}{r^2}+\frac{2M}{r^3}\right.\nn\\
&&\left.+\frac{B^2 m^2}{256}\left[\left(B^2 r^2+8\right) \left(B^4 r^4+8 B^2 r^2+32\right)\right.\right.\nn\\
&&\left.\left.-B^2 r^2\left(B^4 r^4 c^{(6)}_{ll}-B^2 r^2 \left(3 B^2 r^2+16\right) c^{(4)}_{ll}\right.\right.\right.\nn\\
&&\left.\left.\left.+\left(3 B^4 r^4+32 B^2 r^2+96\right) c^{(2)}_{ll}\right)\right]\right\}\,, \\
V_{l\pm 2}&&=f\left[-\frac{B^4 m^2}{256}  r^2 \left(B^4 r^4 c^{(6)}_{l\pm 2l}-B^2 r^2 \left(3 B^2 r^2+16\right) c^{(4)}_{l\pm 2l}\right.\right.\nn\\
&&\left.\left.+\left(3 B^4 r^4+32 B^2 r^2+96\right)   c^{(2)}_{l\pm 2l}\right)\right]\,,\\
V_{l\pm 4}&&=f\left[\frac{B^6 m^2}{256}  r^4 \left(\left(3 B^2 r^2+16\right) c^{(4)}_{l\pm 4l}-B^2 r^2 c^{(6)}_{l\pm 4l}\right)\right]\,,\\
V_{l\pm 6}&&=-f\left[\frac{B^8 m^2}{256}  r^6 c^{(6)}_{l\pm 6l}\right]\,,
\eeq
%
% \end{widetext}
%
where for ease of notation we have suppressed the index $m$ of the Clebsch-Gordan coefficients, but it is understood that the latter depend also on $m$.

This system of equations admits long-lived modes. To find them we can either evolve the system in time (as discussed in Sec.~\ref{sec:time} below) or compute them in the frequency domain. In the frequency domain we consider the following time dependence for the field:
\be
Q_j(r,t)=Q_j(r)e^{-i\omega t}\,.
\ee
Imposing regularity boundary conditions at the horizon and at infinity, Eq.~\eqref{KG_ernst} defines an eigenvalue problem for the complex frequency $\omega=\omega_R+i\omega_I$. The eigenfrequencies are also termed the QNM frequencies and form a discrete spectrum~\cite{Kokkotas:1999bd,Nollert:1999ji,Berti:2009kk}, which generically depends on $m$, $B$ and on the overtone number $n$. Since the presence of the magnetic field breaks the spherical symmetry of the Schwarzschild background, the harmonic index $l$ is not a conserved ``quantum number'' and, for a given $m$, Eq.~\eqref{KG_ernst} effectively describes an \emph{infinite} system of equations where all the eigenfunctions $Q_j$ ($j=0,1,2,...m$) are coupled together.

It is straightforward to show that at the horizon the system decouples and regularity requires purely ingoing waves,
\be\label{hor_ernst}
Q_l(r)\sim e^{-i\omega r_*}\,,\quad r\to r_H\,.
\ee

The behavior at infinity is more intricate since different multipoles are coupled. However this is a difficulty introduced by the spherical coordinates. Expanding the potential~\eqref{potential} at infinity and defining $\rho=r\sin\theta$, we easily see that the asymptotic solution reduces to~\eqref{inf_melvin}. We can then use standard methods for systems of coupled equations (see e.g Refs.~\cite{Rosa:2011my,Pani:2012bp,Brito:2013wya} and the review~\cite{Pani:2013pma}) to find the QNM frequencies. Since the full system~\eqref{Ernst} contains an infinite number of equations, in practice we must truncate the series at some given $L$, i.e. we assume $Q_j\equiv0$ when $j>L$. Convergence is then checked by increasing the truncation order. The results shown have converged to the number of digits displayed and have been obtained with two different methods, a ``direct integration'' and a ``Breit-Wigner'' approach~\cite{Pani:2013pma}.

%%%%%%%%%%%%%%%%%%%%%%%%%%%%%%%%%%%%%%%%%%%%%%%%%%%%%%%%%%%%%%%%%%%
\subsection{Results}
%%%%%%%%%%%%%%%%%%%%%%%%%%%%%%%%%%%%%%%%%%%%%%%%%%%%%%%%%%%%%%%%%%%
%
\begin{table}[t]
%%%%%%%%%%%%%%%%%
% \scriptsize
\centering \caption{Fundamental ($n=0$) QNMs of the Ernst BH solution computed in the frequency domain
for $l=m=1$ and different values of $B$.}
\vskip 12pt
%%%%%%%%%%%%%%%%%%%%%%
\begin{tabular}{@{}ccccccccccccccccccccc@{}}
\hline \hline
$BM$    & $M\omega_R$ & $-M\omega_I$\\
\hline
$0.025$	&	$0.0510$   & $1.2 \cdot 10^{-8}$\\
$0.050$	&	$0.1002$   & $6.7 \cdot 10^{-7}$\\
$0.075$	&	$0.1473$   & $9.1 \cdot 10^{-6}$\\
$0.100$	&	$0.1919$   & $7.0 \cdot 10^{-5}$\\
$0.125$	&	$0.2337$   & $3.7 \cdot 10^{-4}$\\
$0.150$	&	$0.2721$   & $1.4 \cdot 10^{-3}$\\
\hline
\hline
\end{tabular}
\label{tab:Ernst_spectrum1} 
\end{table}

We have performed a detailed numerical analysis of the scalar eigenfrequencies of the Ernst BH as functions of $B$, $m$ and overtone number $n$. Some results are shown in Tables~\ref{tab:Ernst_spectrum1} and~\ref{tab:Ernst_spectrum2}. Even though the background metric is not spherically symmetric, a notion of harmonic index $l$ is still meaningful. In the following we define a mode with given $(l,m)$ as the one corresponding to a set $Q_j$ (with $j=0,1,2,...m$) for which the eigenfunction $Q_l$ is the one with largest relative amplitude. Although this practical definition becomes ambiguous for large values of $B$, we find that such hierarchy in $l$ holds in a large region of the parameter space. For the same reason, the multipolar series in Eq.~\eqref{Ernst} converges rather fast in $L$, even for moderately large values of $B$.

For $BM\ll 1$, the real part behaves approximately as
\be\label{wreal}
\omega_R\sim \left[0.75n+1.2m+0.25l+0.7\right]B\,,
\ee
whereas we infer for the imaginary part a dependence of the form
\be\label{wim}
M\omega_I\sim -\gamma(BM)^{2l+3}\,,
\ee
where $\gamma$ is a numerical coefficient that depends on $n$, $l$ and $m$. For $l=m=1$, we find $\gamma\approx 2.2$ for $n=0$ and $\gamma\approx 9.3$ for $n=1$, respectively.
\begin{table}[t]
%%
% \scriptsize
\centering \caption{Quasinormal modes of the Ernst spacetime computed in the frequency domain for $l=m=1$, $B M=0.1$ and different overtone number $n$.}
\vskip 12pt
\begin{tabular}{@{}ccccccccccccccccccccc@{}}
\hline \hline
$n$ & $M\omega_R$ & $-M\omega_I$\\
\hline
$0$	&	$0.1919$   & $7.0 \cdot 10^{-5}$\\
$1$	&	$0.2674$  & $7.5 \cdot 10^{-4}$\\
$2$	&	$0.3213$  & $2.9 \cdot 10^{-3}$\\
$3$	&	$0.3656$  & $4.5 \cdot 10^{-3}$\\
\hline
\hline
\end{tabular}
\label{tab:Ernst_spectrum2} 
\end{table}
%%%%%%%%%%%%%%%%%%%%%%%%%%%%%%%%%%%

%%%%%%%%%%%%%%%%%%%%%%%%%%%%%%%%%%%%%%%%%%%%%%%%%%%%%%%%%%%%%%%%%%%%%%
\subsection{Relation with previous results in the literature}
%%%%%%%%%%%%%%%%%%%%%%%%%%%%%%%%%%%%%%%%%%%%%%%%%%%%%%%%%%%%%%%%%%%%%%
%
\begin{figure*}[t]
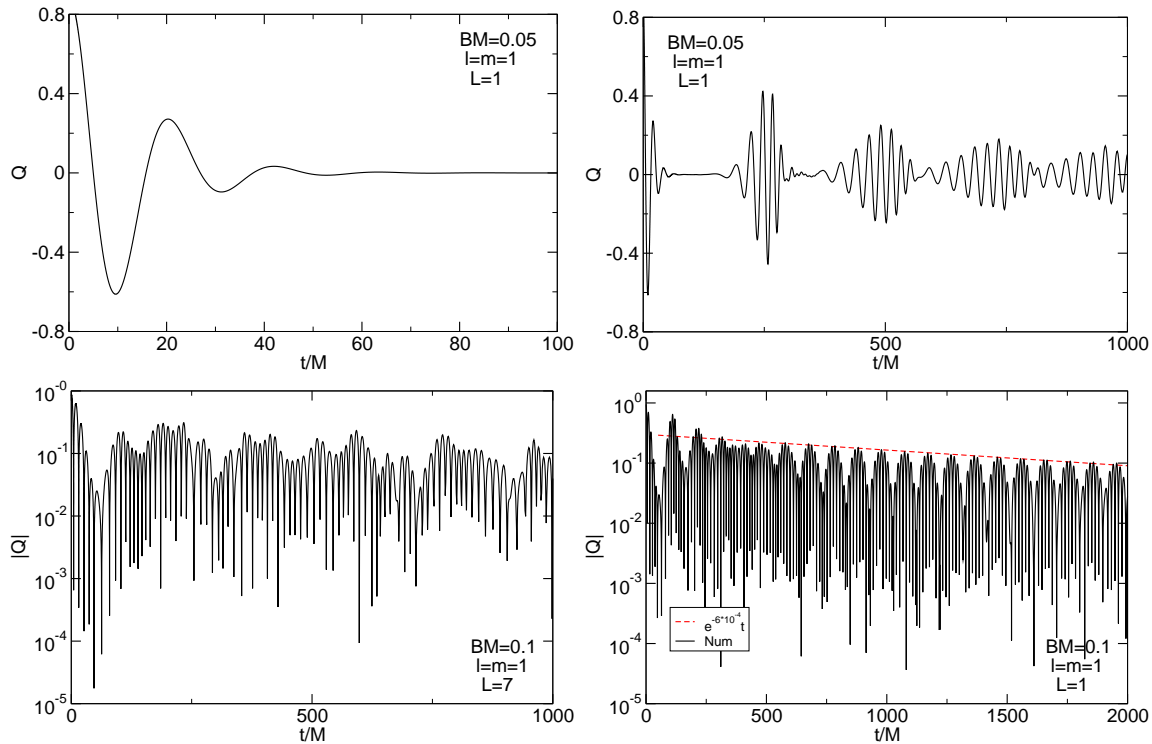

\begin{center}
\begin{tabular}{cc}
\epsfig{file=ringdown_B005.eps,width=7.5cm,angle=0,clip=true}&
\epsfig{file=wave_B005.eps,width=7.5cm,angle=0,clip=true}\\
\epsfig{file=abs_wave_B01.eps,width=7.5cm,angle=0,clip=true}&
\epsfig{file=ana_wave_B01.eps,width=7.5cm,angle=0,clip=true}
\end{tabular}
\caption{Waveforms for a small Gaussian packet $Q_j(0,r)=\delta_{j1}\exp\left[(r_*-r_{c})^2/(2\sigma^2)\right]$ (with $\sigma=6M$ and $r_{c}=6M$ propagating on a Ernst spacetime for different values of $BM$. Top left: BH ringdown at early times for $BM=0.05$ and $l=m=1$. Top right: After a time $t\sim 1/B$, the ``Melvin-like'' modes are excited. Bottom panels: Waveform for $BM=0.1$ and $l=m=1$, truncating the series at $L=7$ (left panel) and $L=1$ (right panel). In the bottom right panel we also show the fit to the decay rate of the scalar field, showing good agreement with the frequency domain analysis.\label{Fig:waveform}}
\end{center}
\end{figure*}
The behavior~\eqref{wreal} and~\eqref{wim} is different from the results of Refs.\cite{Galtsov:1978ag,Konoplya:2007yy,Konoplya:2008hj}. The approximation employed in these works changes the asymptotic behavior at infinity in such a way that the only role of the external magnetic field is to introduce an effective mass, $\mu_{\rm eff}=Bm$, for the scalar field. Consequently, the QNM spectrum was found to be equivalent to that of massive scalar perturbations of a Schwarzschild BH. Massive fields admit quasibound state modes with a hydrogenic spectrum~\cite{Detweiler:1980uk,Dolan:2007mj} 
\begin{equation}
 \omega_R^{\rm mass}\sim\mu_{\rm eff}\,,\qquad M\omega^{\rm mass}_I\sim -(M\mu_{\rm eff})^{4l+6}\,. \label{wmass}
\end{equation}
%%%
While the real part is consistent with the exact behavior~\eqref{wreal}, the scaling of the imaginary part with $B$ is different from Eq.~\eqref{wim}. 

Indeed, solving the full system~\eqref{KG_ernst}, we find that the QNM spectrum has the same qualitative behavior as the modes of a small BH in AdS~\cite{Cardoso:2004hs} or of a BH inside a mirror~\cite{Cardoso:2004nk}.  This is not surprising since the asymptotic behavior plays a crucial role in defining the eigenfrequencies. In fact these frequencies are supported by an effective ``wall'' created by the magnetic field at $r_0\sim 1/B$. In the AdS and in the mirror cases the mirror is given by the AdS radius and by the mirror radius, respectively. We can therefore understand the modes of the Ernst BH as being a small correction to the modes of the Melvin spacetime. The BH event horizon behaves like a perfect absorber~\cite{Thorne:1986iy} and its role is mostly to change one of the boundaries, leading to the slow decay of the field~\footnote{Strictly speaking, since the regular behavior of Eq.~\eqref{inf_melvin} is a damped exponential, such modes could be dubbed as \emph{quasibound} states, in analogy to the case of massive fields~\cite{Dolan:2007mj,Pani:2012vp,Brito:2013wya}. However, in the case of the Ernst solution these are the \emph{only} eigenfrequencies that solve the exact problem and the distinction with the QNMs is irrelevant.}. In the next subsection we shall confirm this qualitative picture using arguments based on the absorption cross section at the horizon and reflections at the asymptotic boundaries [cf. Eq.~\eqref{absor_wIm}].

Therefore, our analysis gives an explicit example of a very natural fact: the eigenvalue spectrum of a given metric is highly sensitive to the asymptotic behavior of the spacetime. Any approximation that changes the boundary conditions might drastically affect the spectrum. Indeed, for the same multipole number $l$ the decay rate of a massive field [cf. Eq.~\eqref{wmass}] can be orders of magnitude smaller than the exact results given in Eq.~\eqref{wim}. Due to this difference, in Sec.~\ref{sec:Ernst_rot} we shall see that, when rotation is turned on, not only the exact modes of the Ernst BH become unstable due to superradiance, but also that the instability time scale can be orders of magnitude \emph{shorter} than that associated to a massive field~\cite{Cardoso:2005vk,Dolan:2007mj} --~and, consequently, the instability is stronger than that discussed in Ref.~\cite{Konoplya:2008hj}.

%%%%%%%%%%%%%%%%%%%%%%%%%%%%%%%%%%%%%%%%%%%%%%%%%%%%%%%%%%%%%%%%%%%%%%
\subsection{Time-domain analysis}\label{sec:time}
%%%%%%%%%%%%%%%%%%%%%%%%%%%%%%%%%%%%%%%%%%%%%%%%%%%%%%%%%%%%%%%%%%%%%%
For completeness, in this section we discuss the results of a time-domain analysis of the system~\eqref{KG_ernst}.
Some examples of waveforms obtained in the time domain are shown in Fig.~\ref{Fig:waveform}. We consider an initial Gaussian wave packet $Q_j(0,r)=\delta_{j1}\exp\left[(r_*-r_{c})^2/(2\sigma^2)\right]$ with $\sigma=6M$ and $r_{c}=6M$, whose time evolution is governed by the system~\eqref{KG_ernst}. The discretization of spatial derivatives is performed using a 2nd order finite difference scheme and integration in time is done with a 4th order accurate Runge-Kutta method.

At early times and for small values of $B$, the waveform is dominated by some ringdown modes~\cite{Berti:2009kk} (top left panel of Fig.~\ref{Fig:waveform}). These ringdown frequencies are \emph{not} given by the modes previously computed, being in fact very similar to the Schwarzschild QNM frequencies~\cite{Berti:2009kk}.
{\it These} numbers are in very good agreement with what was found in Ref.~\cite{Konoplya:2007yy} and are indeed consistent with the fact that for $B M\ll 1$ the ringdown frequencies are almost unaffected by the long-range modification of the potential due to the magnetic field.  

However, we stress that such frequencies do not show up in the frequency domain analysis because they do not satisfy the asymptotic boundary conditions~\eqref{inf_melvin}. Indeed, after a time of order $t\sim 1/B$, the wave is reflected back by the effective wall at $r_0\sim 1/B$ with a smaller amplitude due to the absorption by the event horizon (top right panel of Fig.~\ref{Fig:waveform}). These reflections give rise to the QNM spectrum previously computed in the frequency domain and are related to the Melvin normal modes. Due to the couplings between different multipoles, various frequencies dominate the waveform, making it difficult to extract accurate information about these modes in the time domain.

The fact that the ringdown frequencies at intermediate times are different from the QNMs of the spacetime is another example of an interesting phenomena discussed in detail in Ref.~\cite{Barausse:2014tra} in the context of ``dirty'' BHs.

%%%%%%%%%%%%%%%%%%%%%%%%%%%%%%%%%%%%%%%%%%%%%%%%%%%%%%%%%%%%%%%%%%%%%%%%%%%%%%%%%%%%%%%%%%%%%%%%%%%%%%%%%%%%%%%%
\section{A simple interpretation of the QNMs of a small BH immersed in a confining geometry}\label{sec:model}
%%%%%%%%%%%%%%%%%%%%%%%%%%%%%%%%%%%%%%%%%%%%%%%%%%%%%%%%%%%%%%%%%%%%%%%%%%%%%%%%%%%%%%%%%%%%%%%%%%%%%%%%%%%%%%%%%
The QNM spectra of at least three classes of BH solutions --~namely Ernst BHs immersed in a magnetic field $B\ll1/M$, small AdS BHs (where the AdS radius is much larger than the horizon radius), and BHs confined in a spherical mirror of radius $r_0\gg M$~-- can be all described by a unified picture in terms of a small perfect absorber immersed in a confining box. 
A confining box supports stationary, \emph{normal} modes. Once a small BH is placed inside, one expects that the normal modes will become quasinormal and acquire a small imaginary part, describing absorption -- or amplification -- at the small BH horizon. Thus, it seems that one can separate the two scales -- BH and box size -- and describe quantitatively the system in this way.

Now, normal modes supported by a box have a wavelength comparable to the box size, in other words a frequency $\omega_R\sim1/r_0$. For small BHs, $M/r_0\ll 1$, we then have $M\omega\ll 1$, i.e., we are in the 
low-frequency limit. In this limit, the equation for wave propagation can be solved via matched asymptotics~\cite{Cardoso:2005mh}. The absorption probability at the horizon of a nonrotating BH for a given multipole number $l$ is given by, in this regime~\cite{Cardoso:2005mh}
\beq
\left|\mathcal{A}\right|^2&=& 4\pi \left(\frac{M\omega_R}{2}\right)^{2+2l} \frac{\Gamma^2[1+l+s]\Gamma^2[1+l-s]}{\Gamma^2[1+2l]\Gamma^2[l+3/2]}\nn\\
&\sim&\left({M}/{r_0}\right)^{2l+2}\ll1\, \label{crosssection}
\eeq
where $s=0,2$ for scalar and gravitational fields. Therefore, a wave with initial amplitude $A_0$ is scattered with amplitude $A=A_0\left(1-|\mathcal{A}|^2\right)$ after one interaction with the BH.
Consider now a wave trapped inside the box and undergoing a large number of reflections. After a time $t$ the wave interacted $N=t/r_0$ times with the BH, and its amplitude decreased to
$A=A_0\left(1-|\mathcal{A}|^2\right)^N\sim A_0\left(1-N|\mathcal{A}|^2\right)$. We then get
\be
A(t)=A_0\left(1-t|\mathcal{A}|^2/r_0\right)\,. \label{ampl}
\ee

The net effect of this small absorption at the event horizon is to add a small imaginary part to the frequency, $\omega=\omega_R+i\omega_I$ (with $|\omega_I|\ll\omega_R$). 
In this limit, $A(t)\sim A_0 e^{-|\omega_I| t}\sim A_0(1-|\omega_I| t)$. Comparing the latter equation with Eq.~\eqref{ampl}, we obtain
\be\label{absor_wIm}
M\omega_I\sim-(M/r_0)^{2l+3}\,. 
\ee
This is exactly the scaling that we found numerically, cf. Eq.~\eqref{wim}, when $r_0\sim1/B$. The same behavior is also found in the BH bomb scenario when the mirror is located at $r_0$~\cite{Cardoso:2004nk} and for asymptotically AdS BHs~\cite{Cardoso:2004hs} with the AdS radius being $\sim r_0$. Thus, this provides a consistent, unified description of superradiant instabilities of confined systems.
Our simple model shows that the QNM spectrum of \emph{any} small BH immersed in a confining box of radius $r_0$ scales as $\omega_R\sim1/r_0$ and with a long damping time given by the inverse of Eq.~\eqref{absor_wIm}. This behavior is different from the case of massive perturbations, because the latter can only confine low-frequency modes with $\omega_R\lesssim \mu$ (where $\mu$ is the mass of the field), whereas an ``effective box'' confines radiation of any frequency. From a mathematical viewpoint, this property requires different boundary conditions for the perturbations at infinity.

Finally, our simple model is also useful to capture the essential behavior of:

\noindent{\bf (ii) Gravitational perturbations.} Contrarily to the massive case discussed above, for perturbations confined in an effective box our model predicts that the dependence on the box radius $r_0$ is universal and the same for scalar, vector and tensor fluctuations. Such dependence has in fact been confirmed
recently for the AdS system~\cite{Dias:2013sdc,Cardoso:2013pza}.

\noindent{\bf (iii) Rotating BHs.} Also very relevant to us is that our model can be extended trivially to include rotating BHs. 
When the BH is rotating, low-frequency waves corotating with the BH are amplified by superradiance.
Starobinsky has shown that, at least for moderate spin, the result in Eq.~\eqref{crosssection} still holds with the substitution~\cite{1973ZhETF..64...48S,Staro1,1973ZhETF..65....3S,Staro2}
\be
\omega^{2l+2}\to (\omega-m\Omega_H)\omega^{2l+1}\,,
\ee
where we recall that $\Omega_H$ is the horizon angular velocity. In other words, this intuitive picture immediately predicts that, once rotation is added to the Ernst spacetime, the latter becomes superradiantly unstable. We now turn to show this is indeed the case.

%%%%%%%%%%%%%%%%%%%%%%%%%%%%%%%%%%%%%%%%%%%%%%%%%%%%%%%%%%%%%%%%%%%%%%%%%%%%%%%%%%%%%%%%%%%%%%%%
\section{Superradiant instability of the magnetized Kerr-Newman solution}\label{sec:Ernst_rot}
%%%%%%%%%%%%%%%%%%%%%%%%%%%%%%%%%%%%%%%%%%%%%%%%%%%%%%%%%%%%%%%%%%%%%%%%%%%%%%%%%%%%%%%%%%%%%%%%
After having understood the QNMs of a nonrotating magnetized BH, we now turn our attention to the spinning case. A magnetized rotating BH is a complex object. For example, it was shown by Wald~\cite{Wald:1974np} that when a spinning neutral BH is immersed in a magnetic field it is energetically favorable for it to acquire a charge given by $q=-2\tilde{a} M^2 B$, where $q$ and $\tilde{a}$ correspond to the charge and rotation parameters of the unmagnetized Kerr-Newman solution. This result was established neglecting backreaction effects of the magnetic onto the BH spacetime. Nonetheless, the result was quickly generalized, when Ernst and Wild found the first exact solution of a magnetized Kerr BH~\cite{Ernst:1976:KBH}. This solution was latter shown to suffer from conical singularities at the poles by Hiscock~\cite{Hiscock:1981np}, but he realized that this singularity could be removed by redefining the azimuthal angle $\phi$ (see Appendix~\ref{app:KN}).

Studying perturbations of the full magnetized Kerr-Newman solution (see e.g. Ref.~\cite{Gibbons:2013yq}) is a formidable task. However the problem becomes tractable if we consider an expansion in the rotation parameter $\tilde{a}$. In the following we will consider a slowly-rotating magnetized BH with Wald's charge, to second order in the spin. Note that the slow-rotation approximation of the linear perturbations is fully consistent to second or higher order in $\tilde{a}$, as discussed in Ref.~\cite{Pani:2012vp}.

The Klein-Gordon equation~\eqref{KG} on the slowly-rotating Kerr--Newman background is discussed in Appendix~\ref{app:KN}. The final result reads:
%%%%%%%%%%%%%%
\begin{align}
\sum_l Y_{lm}(\theta,\phi)&\left\{\frac{d^2Q_l(r)}{dr^2_*}+\left[\omega^2-V_{\rm eff}-\tilde{a}m\omega\,W\right]Q_l(r)\right.\nn\\
&\left.+\tilde{a}^2\left[ \sum_{i=0}^4\mathcal{V}_{2i}(r)\cos^{2i}\theta\right] Q_l(r)\right\}=0\,, \label{KG_KN}
\end{align}
%%%
where $V_{\rm eff}=V_{\rm eff}(r,\theta)$ is given by Eq.~\eqref{potential}, the first-order function
%%%
\begin{eqnarray}
 W(r,\theta)&&=\frac{4M^2}{r^3}+\frac{8 M^2 B^2 }{r}-\frac{M^2 B^4}{4}(5r+22M)\nn\\
&&-\frac{M^2 B^4}{4}\cos^2\theta(2+ \cos^2\theta)(r-2M)\,,\nn
\end{eqnarray}
and the second-order radial coefficients $\mathcal{V}_i(r)$ are given in the Appendix.
To separate this equation we use the same technique discussed in Sec.~\ref{sec:Ernst}, leading to an infinite set of radial equations with couplings between different multipoles up to $l\pm 8$.
%%%%%%%%%%%%%%

Defining a tortoise coordinate $dr/dr_*=F$ (where $F$ is a metric variable defined in the Appendix), the purely ingoing wave condition at the horizon reads
\be\label{bchorizon}
Q_l\sim e^{-ik_{H} r_*}\,,\quad r\to r_+\,,
\ee
where $k_H=\omega-m\Omega_H$, $r_+$ is the event horizon radius to second order in $\tilde{a}$ [cf. Eq.~\eqref{rp}], and 
%
%\be
$\Omega_H=-\lim_{r\to r_+} g^{(0)}_{t\phi}/g^{(0)}_{{\phi\phi}}$,
%\ee
%
is the angular velocity at the horizon of locally nonrotating observers, with $g_{\mu\nu}^{(0)}$ being the background metric.

\begin{figure}[hbt]
\begin{center}
%\begin{tabular}{c}
\epsfig{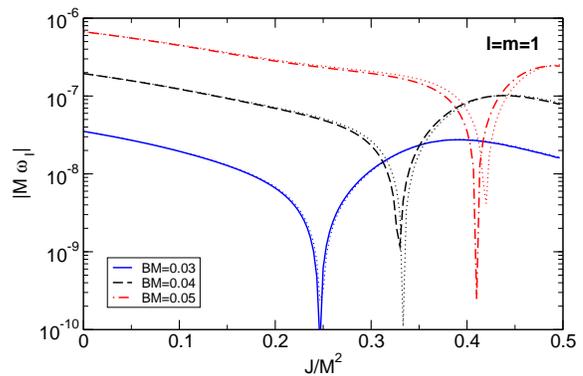}
%\end{tabular}
\caption{Imaginary part of the fundamental modes of a Kerr--Newman--Ernst BH with Wald's charge $q=-2JB$, computed at second order in the rotation and truncating the series at $L=9$, as a function of the BH rotation rate $\tilde{a}=J/M^2$, for $l=m=1$, and different values of the magnetic field. The dotted thinner lines correspond to a magnetized BH without charge. The only effect of the charge is to change the superradiance threshold.\label{Fig:super}}
\end{center}
\end{figure}

We have integrated the eigenvalue problem defined by Eq.~\eqref{KG_KN} numerically. A representative result is shown in Fig.~\ref{Fig:super} where we plot the imaginary part of the fundamental eigenvalue as a function of the BH spin $\tilde{a}\equiv J/M^2$ and for different values of $B$. As discussed in the Appendix, the charge $q$ affects the superradiance threshold. Accordingly, the imaginary part crosses the axis when the superradiant conditions~\eqref{superwald} or~\eqref{supernocharge} are met, for a BH with $q=-2\tilde{a} M^2 B$ or $q=0$, respectively. Although not shown, the real part of the modes depends only mildly on the spin and it is well approximated by Eq.~\eqref{wreal}.

The results shown in Fig.~\ref{Fig:super} are obtained truncating the multipolar series at $L=9$, which guarantees convergence in the entire region of the parameter space under consideration.

In the limit $BM\ll 1$, one can estimate the instability time scale by considering modes~\eqref{wreal} and~\eqref{wim} in the nonrotating case and extrapolating the results to higher values of the spin. The same kind of extrapolation has been done in Refs.~\cite{Pani:2012bp,Brito:2013wya}, where it was found to be sufficiently accurate, for example it captures the onset of the instability and the order of magnitude of the time scale. 
This argument is further supported by the simple model we discussed in Sec.~\ref{sec:model}, and predicts that the imaginary part of the modes scales as
\be
\omega_I M\sim \gamma\left(\tilde{a}m-\frac{2\omega_R r_+}{1+8B^2M^2-16B^4M^4}\right)\left(BM\right)^{2(l+1)}\,.\label{wI_rot}
\ee
Because the time dependence of the perturbation is $\sim e^{i\omega_R t+\omega_I t}$, when the condition~\eqref{superwald} is satisfied $\omega_I>0$ and the perturbation grows exponentially in time.
In other words, as predicted in Sec.~\ref{sec:model}, rotating BHs in Melvin spacetimes are unstable, with an instability time scale given by $1/\omega_I$.

The estimate~\eqref{wI_rot} is in agreement with our numerical results to ${\cal O}(\tilde{a}^2)$. Although our analysis is perturbative in the spin, the results at order $\tilde{a}^2$ are found to be in remarkably good agreement with the exact ones for other systems~\cite{Pani:2012vp,Witek:2012tr}, suggesting that Eq.~\eqref{wI_rot} might be valid beyond its nominal regime of validity. In the next section we take Eq.~\eqref{wI_rot} as an order-of-magnitude estimate to discuss the astrophysical relevance of the superradiant instability triggered by an external magnetic field.

Finally we note that, unlike the case of a massive field, the fundamental mode ($n=0$) does not necessarily have the smallest instability time scale. In fact, the nonrotating results suggest that higher $n$ have larger imaginary parts (see Table~\ref{tab:Ernst_spectrum2}), which translates to a stronger instability in the spinning case. Nonetheless, due to the superradiant condition~\eqref{superwald} and the scaling of $\omega_R$ with $n$ given by Eq.~\eqref{wreal}, only the modes with small $n$ will be superradiant.

%%%%%%%%%%%%%%%%%%%%%%%%%%%%%%%%%%%%%%%%%%%%%%%%%%%%%%%%%%%%%%%%%%%%%%%%%%%%%%%%%%%%%%%%%%%%%%%%
\section{Astrophysical implications of the superradiant instability triggered by magnetic fields}
%%%%%%%%%%%%%%%%%%%%%%%%%%%%%%%%%%%%%%%%%%%%%%%%%%%%%%%%%%%%%%%%%%%%%%%%%%%%%%%%%%%%%%%%%%%%%%%%%  
To measure the strength of a magnetic field in an astrophysical context, we can define the characteristic magnetic field $B_M=1/M$ associated to a spacetime curvature of the same order of the horizon curvature. Restoring physical units, we obtain 
\be\label{magnetic}
B_M\sim 2.4\times 10^{19} \left(\frac{M_{\odot}}{M}\right) {\rm Gauss}\,.
\ee
The strongest magnetic fields around compact objects observed in the Universe are of the order of $10^{13}$--$10^{15} {\rm Gauss}$~\cite{McGill}. In natural units this corresponds to $B/B_M\sim 10^{-6}$--$10^{-4}$. However, $B_M$ is generically much larger than the typical magnetic field believed to be produced by accretion disks surrounding massive BHs. For supermassive BHs with $M\sim 10^9 M_\odot$ a magnetic field $B\sim 10^4 {\rm Gauss}\sim 10^{-6} B_M$ seems to be required to explain the observed luminosity of some active galactic nuclei, assuming a specific model for the interaction between the BH and the accretion disk~\cite{2010arXiv1002.4948P}. Likewise, the typical values of the magnetic field strength near stellar-mass BHs is estimated to be $B\sim 10^8 {\rm Gauss}\sim 10^{-10} B_M$. In other words, the magnetic field near massive BHs typically satisfy $B\ll B_M$. This justifies the small-$B$ estimates given in the previous sections but, on the other hand, it also implies that the superradiant instability time scale would typically be very long.
The purpose of this section is to quantify these statements and to investigate the (superradiant) instability triggered by uniform magnetic fields for astrophysical BHs.

In an astrophysical context our results should be taken with care. The Ernst metric is not asymptotically flat, since it describes a BH immersed in a magnetic field which is supported by some form of ``matter'' at infinity. In a realistic situation, the magnetic field is supported by an accretion disk. The Ernst metric therefore may be a relatively good approximation to the geometry of an astrophysical BH only up to a cutoff distance associated with the matter distribution. In other words, the characteristic length scale $r_0\sim 1/B$ should be smaller than the characteristic distance $r_M$ of the matter distribution around the BH. 
Considering that the accretion disk is concentrated near the innermost stable circular orbit, this would imply that our results can be trusted only when $r_0\lesssim r_M\sim M$, i.e. for $BM\gtrsim 0.1$. As we discussed above, this is a very large value for typical massive BHs. On the other hand, the Ernst metric is more accurate to describe configurations in which the disk extends much beyond the gravitational radius, as is the case in various models. In this case, however, the magnetic field will not be uniform and the matter profile has to be taken into account.  

Nevertheless, and since we wish to make a point of principle, we will use the results obtained in the previous sections for a Kerr BH immersed in a uniform magnetic field to predict interesting astrophysical implications.

\begin{figure}[t]
\begin{center}
%\begin{tabular}{c}
\epsfig{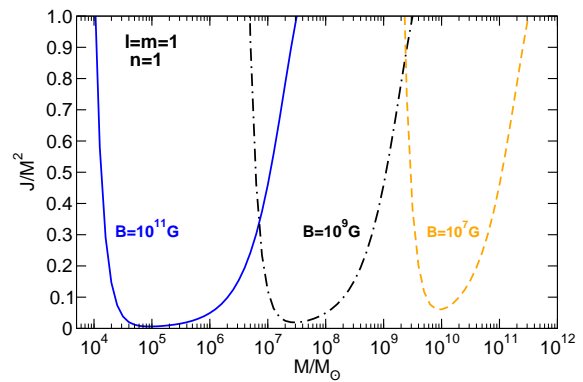}
%\end{tabular}
\caption{Contour plots in the BH Regge plane~\cite{Arvanitaki:2010sy} corresponding to an instability time scale shorter than $\tau_{\rm Salpeter}\sim 4.5\times 10^7{\rm yr}$ for different values of the magnetic field strength $B$
for modes with $l=m=n=1$. BHs lying above each of these curves would be unstable on an observable time scale. The threshold lines are obtained using Eq.~\eqref{wI_rot} in the range $10^{-4}\lesssim BM\lesssim 1$.\label{Fig:Regge}}
\end{center}
\end{figure}
As a result of the superradiant instability, the energy density of the radiation in the region $r\lesssim1/B$ would grow in time at the expense of the BH angular momentum. Therefore, the most likely end state of the instability is a spinning BH with dimensionless spin parameter slightly below the superradiant threshold\footnote{Note that in the case of the full Ernst metric, since radiation cannot escape, the end state is most likely similar to the one in AdS, a rotating BH in equilibrium with the outside radiation~\cite{Cardoso:2006wa}.}. 
This implies an upper limit on the spin of magnetized BHs which depends on the magnetic field, but it is certainly lower than the Kerr bound $\tilde{a}<1$. 
However, this argument remains valid only if the instability extracts the BH angular momentum at higher rate than any possible spin-up effect. For supermassive BHs, the most efficient mechanism to increase the BH spin is prolonged accretion. Therefore, to produce observable effects, the superradiance instability time scale should be shorter than the typical accretion time scale. For accretion at the Eddington rate, the typical time scale is the Salpeter time, $\tau_{\rm Salpeter}\sim 4.5\times 10^7{\rm yr}$.

This type of argument, together with supermassive BH spin measurements (cf. e.g. Refs.~\cite{Brenneman:2011wz,2013Natur.494..449R}), was used to impose stringent constraints on the allowed mass range of axionic~\cite{Arvanitaki:2010sy}, massive vector~\cite{Pani:2012bp,Pani:2012vp} and massive tensor~\cite{Brito:2013wya} fields. Likewise, one could use spin measurements of supermassive BHs to impose constraints on the allowed range of the magnetic field strength. 
In Fig.~\ref{Fig:Regge} we show the spin--mass diagram (so-called BH Regge plane~\cite{Arvanitaki:2010sy}) with contour curves corresponding to an instability time scale $1/\omega_I$ of the order of the Salpeter time. For a given magnetic field $B$, BHs lying above the corresponding threshold curve would be unstable on an observable time scale. 

Spin measurements of supermassive BHs would allow us to locate data points on the Regge plane, thus excluding a whole range of possible magnetic fields. 
Since the contours extend almost up to $J/M^2\sim 0$, one interesting consequence of our results is that essentially \emph{any} observation of a spinning supermassive BH (even with spin as low as $J/M^2\sim 0.1$) would provide some constraint on $B$. However, these observations can possibly exclude only very large values of $B$. For example a putative observation of a supermassive BH with $M\sim 10^9 M_\odot$ and $J/M^2\gtrsim 0.5$ can potentially exclude the range $10^7 {\rm Gauss}\lesssim B\lesssim 10^9 {\rm Gauss}$.

We conclude this section with a note of caution. The threshold lines shown in Fig.~\ref{Fig:Regge} were obtained using Eq.~\eqref{wim} in the range $10^{-4}\lesssim BM\lesssim 1$, but the validity range of Eq.~\eqref{wim} might be smaller. Indeed, a different behavior is expected for large magnetic fields, $BM\gg1$. In the opposite regime, using the magnetized Ernst solution with $BM\sim 10^{-4}$ to approximate a realistic configuration requires the source of the magnetic field to extend at least up to $r_M\sim 10^4M\sim 0.5[M/(10^9 M_\odot)]{\rm pc}$. While we expect that our simplistic analysis can provide the correct order of magnitude for the instability, a more refined study would be needed to assess its validity in the full range of $B$.

%%%%%%%%%%%%%%%%%%%%%%%%%%%%%%%%%%%%%%%%%%%%%%%%%%%%%%%%%
\section{Conclusions \& Extensions}\label{sec:conclusions}
%%%%%%%%%%%%%%%%%%%%%%%%%%%%%%%%%%%%%%%%%%%%%%%%%%%%%%%%%%
The main purpose of this work was to show how strong magnetic fields near spinning BHs can trigger superradiant instabilities and to start exploring the possible implications of such effect. 

To understand this issue, we have computed the normal modes of scalar perturbations of the Melvin spacetime and the QNMs of BHs immersed in a uniform magnetic field. We showed that the magnetic field can confine perturbations leading to long-lived modes, which can trigger superradiant instabilities when the BH spins above a certain threshold. The instability time scale can be orders of magnitude shorter than that associated to the same kind of instabilities triggered by massive fields. In fact, a BH immersed in a uniform magnetic field is very similar to the original BH bomb proposal~\cite{Press:1972zz} and to the case of small BHs in AdS. We provided a simple interpretation of the long-lived modes of such systems in terms of absorption cross section at the horizon and reflection by an effective mirror placed at $r_0\sim 1/B$.

In this work we considered only scalar perturbations.
Due to the presence of the magnetic field, gravitational and electromagnetic perturbations of magnetized BHs are coupled, and even a linear stability analysis is rather involved. Nevertheless, in analogy with the AdS case~\cite{Cardoso:2004hs,Cardoso:2006wa,Dias:2013sdc,Cardoso:2013pza}, we expect the instability of gravito-electromagnetic perturbations of a magnetized Kerr--Newman BH to follow the same scaling as scalar perturbations~\eqref{wI_rot}. This expectation is also supported by the model presented in Sec.~\ref{sec:model}. Since gravitational and electromagnetic perturbations extract energy from the BH more efficiently than a scalar field~\cite{Teukolsky:1974yv} we expect them to trigger a slightly stronger instability. Such problem could be tackled extending the results of Refs.~\cite{Pani:2013ija,Pani:2013wsa}, where the gravito-magnetic modes of a Kerr--Newman BH in vacuum were computed to first order in the spin.

In an astrophysical context, our results should be used with care. The Ernst spacetime is not asymptotically flat and, in a realistic situation, it must be matched with a Minkowski spacetime at large distance. This will add some amount of dissipation which is forbidden in the exact Ernst solution. The validity region of the Ernst metric depends on the extension of the source of the magnetic field. Besides that, in realistic situations the presence of an accretion disk can strongly affect the dynamics of electromagnetic perturbations, for example by quenching growing modes or introducing a cutoff plasma frequency for superradiant photons~\cite{Pani:2013hpa}. 

Nevertheless, we hope our work motivates further studies on the subject. To fully understand the magnitude and end state of the instability, general relativistic magnetohydrodynamic simulations (cf. e.g. Refs.~\cite{2012MNRAS.423.3083M,2013MNRAS.436.3741P}) are necessary. Another related subject that deserves further study are the possible effects of this instability on the Blandford-Znajek process. It would also be interesting to understand if the Meissner effect that affects magnetic fields around highly spinning BHs~\cite{PhysRevD.12.3037,PhysRevD.22.2933,KarasJMP}, and that is still a matter of debate~\cite{Penna:2014aza}, can change the picture in the near-extremal limit.  

Finally, it is possible that a similar superradiant mechanism is at work in rotating stars immersed in strong magnetic fields:
strong fields provide the confinement necessary to grow the superradiant modes, and a putative dissipation at the star
would provide superradiance~\cite{Cardoso:2012zn}. The time scale for energy dissipation in neutron stars is governed by shear viscosity and estimated to be of the order of~\cite{1987ApJ...314..234C}
\be
\tau_{\eta}\sim 10^{9}\left(\frac{10^{14} \,{\rm g}\,{\rm cm}^{-3}}{\rho}\right)^{5/4}\left(\frac{T}{10^9\,{\rm K}}\right)^2\left(\frac{R}{10^6\,{\rm cm}}\right) \,{\rm sec}\,.
\ee
where $\rho$, $T$ and $R$ are the central density of the neutron star, the temperature and the radius, respectively.
By comparison, the time scale for energy dissipation in BHs scales like the light crossing time and is over 14 orders of magnitude smaller
for a stellar-mass BH. Thus, the instability is expected to be of extremely long time scale. Nevertheless, imprints of the (confined) perturbations should appear as new modes of vibration.
%%%%%%%%%%%%%%%%%%%%%%%%%%%%%%%%%%%%%%%%%%%%%%%%%%%%%%%%%%%%%%%%%%%%%%
\begin{acknowledgments}
%%%%%%%%%%%%%%%%%%%%%%%%%%%%%%%%%%%%%%%%%%%%%%%%%%%%%%%%%%%%%%%%%%%%%%
We would like to acknowledge Luis Lehner and Robert Penna for a reading of the manuscript and useful correspondence.
R.B. acknowledges financial support from the FCT-IDPASC program through the Grant No. SFRH/BD/52047/2012, and from the Funda\c c\~ao Calouste Gulbenkian through
the Programa Gulbenkian de Est\' imulo \`a Investiga\c c\~ao Cient\'ifica.
V.C. acknowledges financial support provided under the European Union's FP7 ERC Starting
Grant ``The dynamics of black holes: testing the limits of Einstein's theory''
grant agreement no. DyBHo--256667. 

This research was supported in part by the Perimeter Institute for Theoretical Physics. 
Research at Perimeter Institute is supported by the Government of Canada through 
Industry Canada and by the Province of Ontario through the Ministry of Economic Development 
$\&$ Innovation.
P.P. was supported by the European Community through
the Intra-European Marie Curie Contract No. aStronGR-2011-298297 and by FCT-Portugal through the projects IF/00293/2013 and CERN/FP/123593/2011.
This work was supported by the NRHEP 295189 FP7-PEOPLE-2011-IRSES Grant.
\end{acknowledgments}
%%%%%%%%%%%%%%%%%%%%%%%%%%%%%%%%%%%%%%%%%%%%%%%%%%%%%%%%%%%%%%%%%%%%%%%%%%%%%%

%%%%%%%%%%%%%%%%%%%%%%%%%%%%%%%%%%%%%%%%%%%%%%%%%%%%%%%%%%%%%%%%%%%%%%%%%%%%%%%%%%%%%%%%%%%%%%%%%%%%%%%%%
\appendix
%%%%%%%%%%%%%%%%%%%%%%%%%%%%%%%%%%%%%%%%%%%%%%%%%%%%%%%%%%%%%%%%%%%%%%%%%%%%%%%%%%%%%%
\section{Further details on the  magnetized Kerr-Newman black hole background}\label{app:KN}
%%%%%%%%%%%%%%%%%%%%%%%%%%%%%%%%%%%%%%%%%%%%%%%%%%%%%%%%%%%%%%%%%%%%%%%%%%%%%%%%%%%%%%

The full magnetized Kerr--Newman solution can be found in Refs.~\cite{Ernst:1976:KBH,Diaz:1985xt,Aliev:1989wz,Gibbons:2013yq}.
For $q=-2\tilde{a}M^2 B$ and at second order in the spin, the solution reads
\beq\label{KN_Ernst}
ds^2&=&H\left[-Fdt^2+\Sigma\left(\frac{dr^2}{\Delta}+d\theta^2\right)\right]\nn\\
&+&\frac{A\sin^2\theta}{\Sigma H}\left(H_0 d\phi-\varpi dt\right)^2\,,
\eeq
where $F={\Sigma\Delta}/{A}$, $H_0$ is introduced to remove the conical singularity~\cite{Hiscock:1981np} and 
%
% \begin{widetext}
%
\begin{align}
&\Delta=r^2-2Mr+M^2\tilde{a}^2+q^2\,,\\
&\Sigma=r^2+\tilde{a}^2M^2\cos^2\theta\,,\\
&A=r^4+M^2 r \tilde{a}^2 \left[\sin ^2\theta (2 M-r)+2 r\right]\,,\\
&H=1+\frac{1}{2}B^2 r^2 \sin ^2\theta+\frac{1}{16} B^4 r^4 \sin ^4\theta \nn\\
&+\tilde{a}^2\left[\frac{1}{8} B^6 M^4 r^2 \sin ^2 2\theta
+\frac{1}{8} B^4 M^2\left(2 M r \sin ^6\theta\right.\right.\nn\\
&\left.\left.+2
   M \cos ^2\theta \left(M \cos ^4\theta+2 \cos ^2\theta(M-2 r)+9 M+8 r\right)\right.\right. \nn\\
&\left.\left.-8 M r+r^2 \sin ^4\theta\right)+\frac{B^2 M^2}{2 r}
 \sin ^2\theta (r-M(7+\cos 2\theta))\right]\,,\\
&H_0\equiv H(r,\theta=0)=1+3 B^4 M^4 \tilde{a}^2\,,\\
&\varpi=\frac{M^2 \tilde{a}}{64 r^3} \left[-B^4 r^3 (12 \cos 2\theta+\cos 4 \theta) (r-2 M)\right.\nn\\
&\left.+B^2 r^2 \left(256-B^2 r (154 M+51   r)\right)+128\right]\,.
\end{align}
%
% \end{widetext}
This solution reduces to the Ernst metric~\eqref{Ernst} when $\tilde{a}=0$. To second order in $\tilde{a}$, the event horizon is located at
\be
r_+=2M-\tilde{a}^2\left(\frac{M}{2}+2B^2M^3\right)\,, \label{rp}
\ee
and the vector potential of the magnetic field is given by
\be
A=\Phi_0 dt+\Phi_3\left(H_0 d\phi-\varpi dt\right)\,.
\ee
The explicit form of the functions $\Phi_0$ and $\Phi_3$ is not important here, so we refer the reader to Ref.~\cite{Gibbons:2013yq}.

Interestingly, these solutions incorporate Wald's result for the charge induction~\cite{Wald:1974np} in the small-$B$ limit. This allows us to understand the Wald's charge as being the one needed to have a vanishing total electric charge at infinity. Indeed the total physical charge of the solution is given by~\cite{Aliev:1988wy,Gibbons:2013yq}
\be\label{charge}
Q=q\left(1-\frac{1}{4}q^2B^2\right)+2\tilde{a} M^2 B\,.
\ee
Due to the vacuum polarization and accretion of particles of opposite charge, BHs have a tendency to quickly lose their charge~\cite{Gibbons:1975kk}. In order to be neutral, a BH must then satisfy $q\left(1-\frac{1}{4}q^2B^2\right)=-2\tilde{a} M^2 B$.
Solving for $q$ and expanding in the small-$B$ limit we find
\be
q_{\rm neutral}/M=-2\tilde{a}BM+\mathcal{O}\left[\tilde{a}^3(BM)^5\right]\,.
\ee
The result above reduces to Wald's results to first order in $BM$ and also in the small-rotating limit.

Note that $q$ and $\tilde{a}$ do not have a direct physical meaning for the exact geometry of the magnetized BH. The conserved electric charge of the magnetized BH is given by $Q$~\eqref{charge}, while the true conserved angular momentum of the exact magnetized BH solutions can be evaluated from thermodynamic considerations, as it was done in Ref.~\cite{Gibbons:2013dna}. Although this quantity can be quite complicated, expanding in the small-$\tilde{a}$ limit and considering a BH with Wald's charge, one recovers the standard relation for the angular momentum of a Kerr BH,
\be
J=\tilde{a}M^2+\mathcal{O}\left(\tilde{a}^3\right)\,.
\ee

For a BH with charge $q=-2\tilde{a}M^2B+\mathcal{O}\left(\tilde{a}^3\right)$, the horizon's angular velocity $\Omega_H$ is given by
\be\label{angular_vel}
\Omega_H=\frac{\tilde{a}}{4M}+2\tilde{a}M B^2\left(1-2B^2M^2\right)+\mathcal{O}\left(\tilde{a}^3\right)\,.
\ee
Note that $\Omega_H$ is slightly different from the case of a magnetized BH with $q=0$. Indeed, when $q\neq 0$ a charged BH has a gyromagnetic ratio $q/M$~\cite{Carter:1968rr}, so it can acquire an angular momentum when immersed in a uniform magnetic field. The extra term proportional to $B$ in~\eqref{angular_vel} is related to this effect. This can be seen by computing $\Omega_H$ for a BH with $\tilde{a}=0$,
\beq\label{angular_charged}
\Omega^{(\tilde{a}=0)}_H&=&-\frac{8 qB \left[B^2 \left(q^2-4 M \sqrt{M^2-q^2}-4 M^2\right)+4\right]}{\left(B^4 q^4+24 B^2 q^2+16\right)
   \left(\sqrt{M^2-q^2}+M\right)}\nn\\
	&\sim&-\frac{qB}{M}+2B^3M q+\mathcal{O}\left(q^3M^3\right)\,,
\eeq
where in the last step we linearized in $q$. Taking $q=q_{\rm neutral}$ we get the extra term proportional to $B$ in~\eqref{angular_vel}.

Due to the boundary condition~\eqref{bchorizon}, superradiant scattering is possible whenever $\omega_R<m\Omega_H$~\cite{Teukolsky:1974yv} or (to second order in rotation):
\be\label{superwald}
\tilde{a}>\frac{4M\omega_R}{m\left(1+8B^2M^2-16B^4M^4\right)}\,,
\ee
where $\omega_R$ is the real part of the mode frequency given approximately by~\eqref{wreal}. The effect of the charge induced by the magnetic field is to change the superradiant threshold which, for a BH with $q=0$, is given by
\be\label{supernocharge}
\tilde{a}>\frac{4M\omega_R}{m}\,.
\ee

Finally, after some algebra the Klein-Gordon equation~\eqref{KG} in the background~\eqref{KN_Ernst} reduces to Eq.~\eqref{KG_KN} in the main text, where the coefficients are given by
\begin{widetext}
\begin{align}
\mathcal{V}_0&=\frac{3 B^{12} M^4 m^2}{128}  (r-2M) r^5+\frac{1}{64} B^{10} M^4 m^2 r^3 (23 r-48M)-\frac{B^8M^2 m^2}{128}  \left(r^4-968 M^4+136 M^3 r-280 M^2 r^2-14 M r^3\right) \nn\\
&-\frac{B^6 M^2
   m^2 \left(544 M^3+48 M^2 r-20 M r^2+r^3\right)}{16 r}+\frac{B^4 M^3m^2 (9 r^2-46 M^2+10 M r)}{2 r^3}\nn\\
&+\frac{B^2 M^2\left(r \left(-4 l (l+1)M^2+m^2 r
   (r+4M)+8M^2\right)-24M^3\right)}{r^5}\nn\\
&	+\frac{M^2}{r^5} \left[l(l+1)(r-4M)+r \left(m^2-(r-2M) r \omega ^2-1\right)+12M\right]-24M^4\,,\\
\mathcal{V}_2&=-\frac{9}{128} B^{12} M^4 m^2 (r-2M) r^5+\frac{1}{64} B^{10} M^4 m^2 (104M-49 r) r^3\nn\\
&+\frac{B^8 M^2 m^2}{256}  \left(-704 M^4+1744 M^3 r-424 M^2 r^2-76 M r^3+5 r^4\right)+\frac{B^6 M^2 m^2 (r+2M) ((r-36M) r+84M^2)}{16 r}
\nn\\
&-\frac{B^4 M^2 m^2 \left[8 M^3 + r (-76 M^2 + 3 r (8 M + r))\right]}{8 r^3}-\frac{B^2M^2 m^2
   (r-2M)}{r^3}+\frac{(r-2M) M^2\omega ^2}{r^3}\,,\\
\mathcal{V}_4&=\frac{9}{128} B^{12}M^4 m^2 (r-2M) r^5+\frac{1}{64} B^{10}M^4 m^2 r^3 (29 r-64M)\nn\\
&-\frac{1}{256} B^8 M^2 m^2 \left[288 M^4+r \left(336 M^3+r \left(3 r (r-20 M)-56 M^2\right)\right)\right]\nn\\
&+\frac{B^6M^2
   m^2 \left[48 M^3+r (r-4 M) (12 M+r)\right]}{16 r}+\frac{3 B^4M^2 m^2 (r-2M)^2}{8 r^2}\,,\\
\mathcal{V}_6&=-\frac{3}{128} B^{12}M^4 m^2 (r-2M) r^5+\frac{1}{64} B^{10}M^4 m^2 (8M-3 r) r^3\nn\\
&-\frac{1}{256} B^8 M^2 m^2 \left[-64 M^4+80 M^3 r-40 M^2 r^2+4 M r^3+r^4\right]-\frac{B^6 M^2 m^2
   (r-2M)^3}{16 r}\,,\\
\mathcal{V}_8&=\frac{1}{256}  M^2B^8 m^2 (r-2M)^4\,.
\end{align}
\end{widetext}
% 

%%%%%%%%%%%%%%%%%%%%%%%
\bibliography{ref}  
%%%%%%%%%%%%%%%%%%%%%%%%%%

\end{document}